\documentclass{aa}
\usepackage{graphicx}
\usepackage{hyperref}
\usepackage{lscape}
\usepackage{natbib}
\usepackage[scaled]{helvet}
\usepackage[varg]{txfonts}
\usepackage{url}
\usepackage{breakurl}
\usepackage{xspace}
\bibpunct{(}{)}{;}{a}{}{,}
\newcounter{Rco}
\newcommand{\ionw}[3]{\mbox{\ion{#1}{#2}~$\lambda\,#3\,\mathrm{\AA}$}\xspace}
\newcommand{\ionww}[3]{\mbox{\ion{#1}{#2}~$\lambda\lambda\,#3\,\mathrm{\AA}$}\xspace}
\newcommand{\Ionst}[1]{\setcounter{Rco}{#1}\Roman{Rco}}
\newcommand{\Ion}[2]{\mbox{#1\,{\scriptsize\Ionst{#2}}}}
\newcommand{\Ionw}[3]{\mbox{#1\,{\scriptsize\Ionst{#2}}~$\lambda\,#3$\,\AA}\xspace}
\newcommand{\Ionwf}[3]{\mbox{[#1\,{\scriptsize\Ionst{#2}}]~$\lambda\,#3$\,\AA}\xspace}
\newcommand{\Ionww}[3]{\mbox{#1\,{\scriptsize\Ionst{#2}}~$\lambda\lambda\,#3$\,\AA}\xspace}
\newcommand{\Jonw}[3]{\mbox{\ion{#1}{#2}~$\lambda\,#3$\,\AA}\xspace}
\newcommand{\Jonww}[3]{\mbox{\ion{#1}{#2}~$\lambda\lambda\,#3$\,\AA}\xspace}
\newcommand{\ea}{et al\@. }
\newcommand{\logg}{\mbox{$\log g$}\xspace}
\newcommand{\loggw}[1]{\mbox{$\log g\hspace{-0.5mm} =\hspace{-0.5mm}  #1$}}
\newcommand{\kK}{\mathrm{kK}}
\newcommand{\ab}[1]{\mbox{Fig.\,\ref{#1}}}
\newcommand{\sA}[1]{\mbox{(Fig.\,\ref{#1})}}
\newcommand{\ratio}[2]{\mbox{$n_{\rm #1}/n_{\rm #2}$}}
\newcommand{\ratiow}[3]{\mbox{$n_{\rm #1}/n_{\rm #2}\hspace{-0.5mm} = \hspace{-0.5mm} #3$}}
\newcommand{\se}[1]{\mbox{Sect.\,\ref{#1}}}
\newcommand{\sga}{\raisebox{-0.10em}{$\stackrel{>}{{\mbox{\tiny $\sim$}}}$}}
\newcommand{\sK}[1]{\mbox{(Sect.\,\ref{#1})}}
\newcommand{\sla}{\raisebox{-0.10em}{$\stackrel{<}{{\mbox{\tiny $\sim$}}}$}}
\newcommand{\spm}{\mbox{\raisebox{0.20em}{{\tiny \hspace{0.2mm}\mbox{$\pm$}\hspace{0.2mm}}}}}
\newcommand{\ta}[1]{\mbox{Tab.\,\ref{#1}}}
\newcommand{\sT}[1]{\mbox{(Tab.\,\ref{#1})}}
\newcommand{\Teff}{\mbox{$T_\mathrm{eff}$}\xspace}
\newcommand{\Teffw}[1]{\mbox{$\Teff\hspace{-0.5mm} =\hspace{-0.5mm} #1 \,\mathrm{K}$}}
\newcommand{\ebv}{\mbox{$E_\mathrm{B-V}$}}
\newcommand{\ebvw}[1]{\mbox{$\ebv\hspace{-0.5mm} =\hspace{-0.5mm} #1$}}
\newcommand{\deh}{\mbox{$N_\ion{D}{i}$}}
\newcommand{\nh}{\mbox{$N_\ion{H}{i}$}}
\newcommand{\nhw}[1]{\mbox{$\nh\hspace{-0.5mm} =\hspace{-0.5mm} #1\, \mathrm{cm}^{-2}$}}
\newcommand{\vrad}{\mbox{$v_\mathrm{rad}$}}
\newcommand{\vradw}[1]{\mbox{$\vrad = \hspace{-0.5mm} #1\, \mathrm{km\,sec}^{-1}$}}
\newcommand{\Msol}{$M_\odot$\xspace}
\newcommand{\lmspr}{\hbox{}\hspace{+1.2cm}}
\newcommand{\mmspr}{\hbox{}\hspace{+0.7cm}}
\newcommand{\smspr}{\hbox{}\hspace{+2.5mm}}
\newcommand{\smspl}{\hbox{}\hspace{+0.1mm}}
\newcommand{\gb}{\object{G191$-$B2B}\xspace}
\newcommand{\re}{\object{RE\,0503$-$289}\xspace}
\newcommand{\wdn}{\object{WD\,0111$+$002}\xspace}
\newcommand{\pgs}{\object{PG\,1707$+$427}\xspace}
\newcommand{\pgn}{\object{PG\,0109$+$111}\xspace}
\newcommand{\bd}{BD$-22\degr 3467$\xspace}
\newcommand{\hd}{HD\,127493\xspace}
\newcommand{\hz}{HZ\,44, \xspace}
\newcommand{\feige}{Feige\,46\xspace}
\newcommand{\lsiv}{LS\,IV$-14\degr 116$\xspace}
\begin{document}

\title{Stellar laboratories }
\subtitle{X. New \ion{Cu}{iv -- vii} oscillator strengths and \\ the first detection of copper and indium in hot white dwarfs
           \thanks
           {Based on observations with the NASA/ESA Hubble Space Telescope, obtained at the Space Telescope Science 
            Institute, which is operated by the Association of Universities for Research in Astronomy, Inc., under 
            NASA contract NAS5-26666.
           }\fnmsep
           \thanks
           {Based on observations made with the NASA-CNES-CSA Far Ultraviolet Spectroscopic Explorer.
           }\fnmsep
           \thanks
           {Tables \ref{tab:cuiv:loggf} to \ref{tab:cuvii:loggf} are only available via the
            German Astrophysical Virtual Observatory (GAVO) service TOSS (\url{http://dc.g-vo.org/TOSS}).
           }
         }

\titlerunning{Stellar laboratories: New \ion{Cu}{iv -- vii} oscillator strengths}

\author{T\@. Rauch\inst{1}
        \and
        S\@. Gamrath\inst{2}
        \and
        P\@. Quinet\inst{2,3}
        \and
        M\@. Demleitner\inst{4}
        \and
        M\@. Kn\"orzer\inst{1}
        \and
        K\@. Werner\inst{1}
        \and
        J\@. W\@. Kruk\inst{5}
        }

\institute{Institute for Astronomy and Astrophysics,
           Kepler Center for Astro and Particle Physics,
           Eberhard Karls University,
           Sand 1,
           72076 T\"ubingen,
           Germany \\
           \email{rauch@astro.uni-tuebingen.de}
           \and
           Physique Atomique et Astrophysique, Universit\'e de Mons -- UMONS, 7000 Mons, Belgium
           \and
           IPNAS, Universit\'e de Li\`ege, Sart Tilman, 4000 Li\`ege, Belgium
           \and
           Astronomisches Rechen-Institut (ARI), Centre for Astronomy of Heidelberg University, 
                M\"onchhofstra\ss e 12-14, 69120 Heidelberg, Germany
           \and
           NASA Goddard Space Flight Center, Greenbelt, MD\,20771, USA}

\date{Received 3 September 2019; accepted 23 March 2020}

\abstract {Accurate atomic data is an essential ingredient for the calculation of reliable
           non-local thermodynamic equilibrium (NLTE) model atmospheres that are mandatory
           for the spectral analysis of hot stars.
          }
          {We aim to search for and identify for the first time spectral lines of copper (atomic number $Z = 29$) and
           indium ($Z = 49$) in hot 
           white dwarf (WD) stars and to subsequently determine their photospheric abundances.
          }
          {Oscillator strengths of \ion{Cu}{iv - vii} were calculated to include radiative and 
           collisional bound-bound transitions of Cu in our NLTE model-atmosphere calculations.
           Oscillator strengths of \ion{In}{iv - vi} were compiled from the literature.
          }
          {We newly identified
            1 \ion{Cu}{iv}, 
           51 \ion{Cu}{v},
            2 \ion{Cu}{vi}, and
            5 \ion{In}{v}
           lines in the ultraviolet (UV) spectrum of DO-type WD \re.
           We determined the photospheric abundances of 
           $9.3 \times 10^{-5}$ (mass fraction, 132 times solar)
           and
           $3.0 \times 10^{-5}$ (56\,600 times solar),
           respectively;
           we also found Cu overabundances in the DA-type WD \gb ($6.3 \times 10^{-6}$,
           9 times solar).
          }
          {All identified \ion{Cu}{iv-vi} lines in the UV spectrum of \re 
           were simultaneously well reproduced with our newly calculated oscillator strengths.
           With the detection of Cu and In in \re, the total number of trans-iron elements ($Z > 28$)
           in this extraordinary WD reaches an unprecedented number of 18.
          }

\keywords{atomic data --
          line: identification --
          stars: abundances --
          stars: individual: \gb\ --
          stars: individual: \re\ --
          virtual observatory tools
         }

\maketitle

\section{Introduction}
\label{sect:intro}

\begin{figure}
   \resizebox{\hsize}{!}{\includegraphics{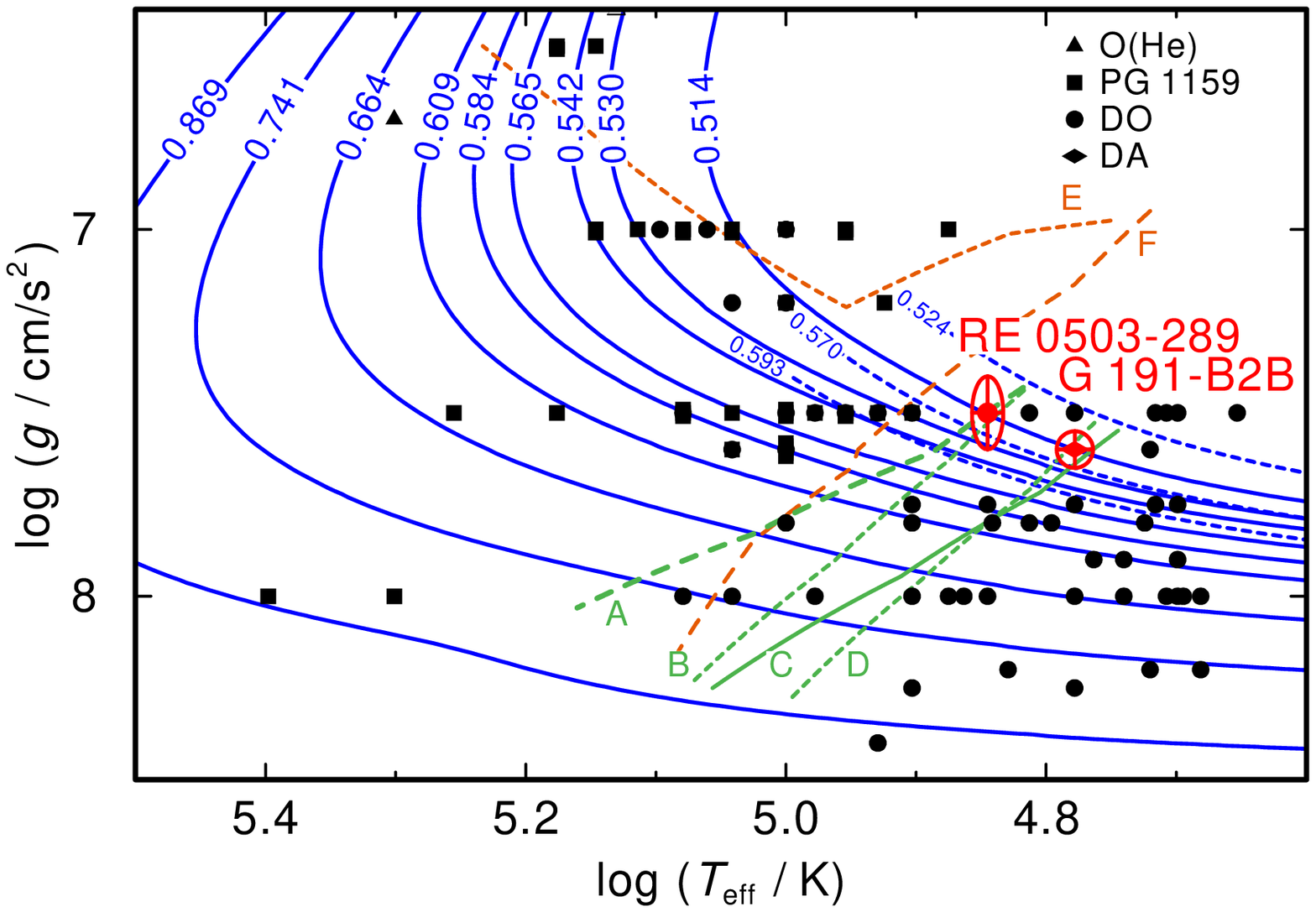}}
   \caption[dummy]
           {Location of \re (with its error ellipse) and related objects
            \citep{huegelmeyeretal2006,kepleretal2016,reindletal2014ohe,reindletal2014do,wernerherwig2006}
            in the \mbox{$\log T_\mathrm{eff}$\,--\,\logg} plane\footnotemark.
            Evolutionary tracks for H-deficient \citep[full lines]{althausetal2009} and
            H-rich WDs \citep[dashed lines]{millerbertolami2016}
            labeled with their respective masses in $M_\odot$ are plotted for comparison.
            ``Wind limits'' of \citet[their Figs.\,6 and 13,][digitized with Dexter\footnotemark]{unglaubbues2000} 
            are shown. The lines indicate for
            H-deficient stars
            A: so-called PG\,1159 wind limit calculated with $\dot{M} = 1.29\times 10^{-15}L^{1.86}$;
            B and C: photospheric carbon content is reduced by factors of 0.5 and 0.1, respectively;
            D: no PG\,1159 star is known to the right of this line,
            and for H-rich stars
            E: wind limit; and 
            F: He/H = 10$^-3$ (by number).
           }
   \label{fig:evol}
\end{figure}

The hydrogen-deficient (DO-type) white dwarf (WD) \re \citep[WD\,0501-289;][]{mccooksion1999,mccooksion1999cat}
is located in the $\log T_\mathrm{eff}$\,--\,\logg\ plane
\citep[Fig.\,\ref{fig:evol}, \Teffw{70\,000 \pm 2000} and $\log\,(g\,/\,\mathrm{cm\,s^{-2}}) = 7.5 \pm 0.1$;][]{rauchetal2016kr}
in the transition 
zone between those post-asymptotic giant branch (post-AGB) stars with still high 
luminosity and a stellar wind still strong enough to provide a chemically homogeneous 
photosphere and the WD cooling sequence where gravity
efficiently pulls metals down and out of the photosphere. In this region the interplay of
radiative levitation and gravitational settling is responsible for stratification, i.e.,
metals float up and, thus strong abundance enhancements are observed.  Trans-iron
elements (TIEs), with their many spectral lines due to their only partially filled electron
shells, are especially involved in this process. Thus, although unexpected, it was not surprising
that \citet{werneretal2012} discovered lines of ten TIEs in the ultraviolet (UV) spectrum
of \re. Due to the lack of atomic data for TIEs at that time, it was only possible to
determine the Kr and Xe abundances that are about 450 and 3800 times solar, respectively.

Since 2012 we have calculated oscillator strengths of several
TIEs (Table\,\ref{tab:tee}) and successfully identified their lines in the spectra of
\re,
\gb \citep[WD\,0501+527;][see additional references in Table\,\ref{tab:tee}]{mccooksion1999,mccooksion1999cat},
\wdn, \pgn, and \pgs \citep{hoyeretal2017}. To search for Cu and In lines in the UV
spectrum of \re and \gb, we calculated new
oscillator strengths for Cu and compiled In data from
the literature. In Sect.\,\ref{sect:observation}, we briefly describe the available UV
observations. Our model-atmosphere code and the considered atomic data is introduced in
Sect.\,\ref{sect:models}. We summarize our results and conclude in Sect.\,\ref{sect:results}.

\addtocounter{footnote}{-1}
\footnotetext{cf. \url{http://www.astro.physik.uni-potsdam.de/~nreindl/He.html} for stellar parameters}
\addtocounter{footnote}{1}
\footnotetext{\url{http://dc.zah.uni-heidelberg.de/sdexter}}

\section{Observations}
\label{sect:observation}

For our spectral analysis, we used UV spectra obtained with 
the Far Ultraviolet Spectroscopic Explorer 
(FUSE, $910\,\mathrm{\AA} < \lambda < 1190\,\mathrm{\AA}$, resolving power $R \approx 20\,000$) and
the Hubble Space Telescope / Space Telescope Imaging Spectrograph 
(HST/STIS, $1144\,\mathrm{\AA} < \lambda < 1709\,\mathrm{\AA}$ with
$R=2.3 \times 10^5$ for \gb
and
$R \approx 45\,800$ for \re).
All spectra are described in detail in \citet{rauchetal2013} and \citet{hoyeretal2017}.

The observed spectra shown here 
(FUSE for $\lambda \le 1188\,\mathrm{\AA}$, STIS otherwise)
were shifted to rest wavelengths, using
$v_\mathrm{rad} = 24.56\,\mathrm{km\,s^{-1}}$ for \gb \citep{lemoineetal2002} and 
                 $25.8\,\mathrm{km\,s^{-1}}$ for \re \citep{hoyeretal2017}.
To compare them with our synthetic spectra, they were convolved with Gaussians to 
simulate the respective instrument resolving power.

\section{Model atmospheres and atomic data}
\label{sect:models}

For the precise spectral analysis of hot stars, model atmospheres that consider deviations from
the local thermodynamical equilibrium (LTE) are mandatory. The T\"ubingen non-local thermodynamic
equilibrium (NLTE) model-atmosphere package 
\citep[TMAP,][]{tmap2012} can calculate such models that
it assumes radiative and hydrostatic equilibrium and  plane-parallel geometry.
The T\"ubingen Model Atom Database
\citep[TMAD,][]{rauchdeetjen2003} provides
the model atoms for elements with atomic number below 20.
TMAD has been constructed as part of the T\"ubingen contribution to the 
German Astrophysical Virtual Observatory 
(GAVO).

The ionization stages of \ion{Cu}{iv-vii} and \ion{In}{iv-vi} were represented in the model atoms
using so-called super levels and super lines. These were calculated via a statistical approach by our 
Iron Opacity and Interface 
\citep[IrOnIc,][]{rauchdeetjen2003,muellerringatPhD2013}.
To enable IrOnIc to read our new Cu and In data, we transferred it into Kurucz-formatted 
files \citep[cf.][]{rauchetal2015ga}. 
The statistics of our Cu and In model atoms are listed in Table\,\ref{tab:cuinironic}.

\begin{table}\centering
\caption{Statistics of the \ion{Cu}{iv - vii},
         atomic levels and line transitions from
         Tables\,\ref{tab:cuiv:loggf} - \ref{tab:cuvii:loggf}
         and of \ion{In}{iv - vi} compiled from the literature (Sect.\,\ref{sect:results}). 
         The super levels and lines are used in our model-atmosphere calculations.
        } 
\label{tab:cuinironic}
\begin{tabular}{r@{\,}lcccc}
\hline
\hline
\multicolumn{2}{c}{Ion}        & Atomic levels & Lines & Super levels & Super lines \\
\hline
Cu&{\sc iv}                    &           297 &  8785 &            7 &          15 \\
  &{\sc v}                     &           247 &  5456 &            7 &          16 \\
  &{\sc vi}                    &           254 &  3797 &            7 &           9 \\
  &{\sc vii}                   &           217 &  2253 &              &             \\
In&{\sc iv}                    &           231 &   564 &            7 &          15 \\
  &{\sc v}                     &           396 &   919 &            7 &          16 \\
  &{\sc vi}                    &           167 &   176 &            8 &           9 \\
\hline
\end{tabular}
\end{table}

\begin{figure}
   \resizebox{\hsize}{!}{\includegraphics{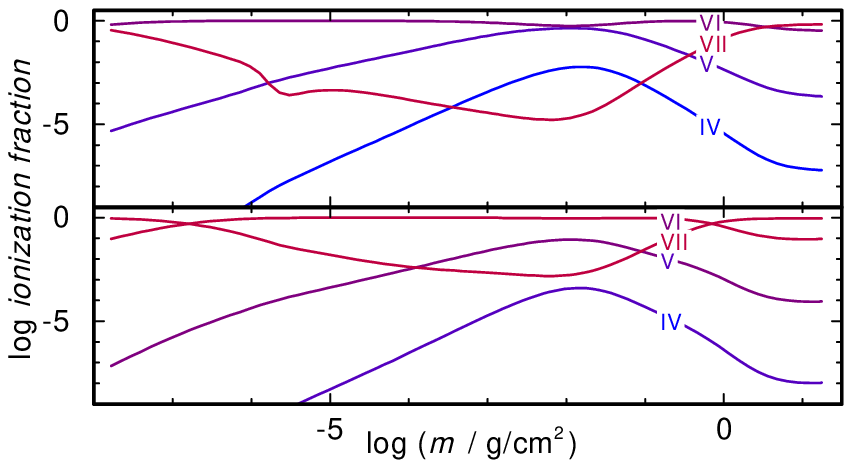}}
    \caption{Ionization fractions of Cu (\emph{top} panel) and 
             In (\emph{bottom})  in our \re model.
             $m$ is the column mass, measured from the outer boundary of our model atmospheres.
            }
   \label{fig:ion_cuin_re}
\end{figure}

Our models consider opacities 
of 
He,           
C,            
N,            
O,            
Al,           
Si,           
P,            
S,            
Ca,           
Sc,           
Ti,           
V,            
Cr,           
Mn,           
Fe,           
Co,           
Ni,           
Cu,
Zn,           
Ga,           
Ge,           
As,           
Se,           
Kr
Sr,           
Zr,           
Mo,           
Sn,           
In,
Te,           
I,           
Xe, and
Ba. Details about the models are given in \citet{rauchetal2016kr}.
\ion{Cu}{v-vi} and \ion{In}{vi} are the
dominant ionization stages in the line-forming region ($-4 \la \log\,m \la 0.5$)
of the photosphere of \re (Fig.\,\ref{fig:ion_cuin_re}).

The calculation of the Cu and In absorption cross sections is performed according to \citet{rauchdeetjen2003}.
For the collisional bound-bound transitions we use the van Regemorter formula \citep{vanregemorter1962} 
for known f-values and an approximate formula for unknown f-values.
The quadratic Stark effect is considered for radiative bound-bound transitions by an approximate formula 
given by \citep{cowley1970,cowley1971}.
The \citet{seaton1962} formula is employed to calculate collisional and radiative bound-free
cross sections with a hydrogen-like threshold value.

For Cu and In (and all other elements), level dissolution (pressure ionization) is considered
following \citet{hummermihalas1988} and \citet{hubenyetal1994}.

\paragraph{Available atomic data of \ion{Cu}{iv-vii} ions.}

The main source of atomic data related to the \ion{Cu}{iv-vii} spectra is
the paper published by 
\citet{sugarmusgrove1990} 
in which the available
experimental energy levels of the copper atom, in all stages of
ionization, have been compiled together with ionization energies,
either experimental or theoretical, experimental Landé g-factors, and
leading components of calculated eigenvectors. This compilation is
still being used as the standard reference database for the copper ions of
interest at the National Institute of Standards and Technology
\citep[NIST,][]{kramidaetal2019}.  
More precisely, in Sugar and Musgrove and NIST
compilations for \ion{Cu}{iv}, experimental values are reported for 187
levels of the 
3d$^8$, 
3d$^7$4s, 
3d$^7$5s, 
3d$^7$4d, and 
3d$^6$4s$^2$ 
even-parity configurations and 110 levels of the 
3d$^7$4p 
odd-parity
configuration. These data are based on earlier analyses by 
\citet{schroedervankleef1970}, 
\citet{meinders1976}, and 
\citet{meindersuijlings1980}
who
observed the copper spectrum using a sliding spark light source.  In
the case of \ion{Cu}{v}, the first work was reported by 
\citet{vankleefetal1975} 
who found a few energy levels in 
3d$^7$ and 
3d$^6$4p
configurations. This work was completed by 
\citet{vankleefetal1976} 
who identified many 
3d$^7$ - 3d$^6$4p and 
3d$^6$4s - 3d$^6$4p 
lines, allowing them to
classify all the levels of the 
3d$^7$ 
configuration, as well as 53 of the
63 levels of 
3d$^6$4s 
and 175 of the 180 levels of 
3d$^6$4p.  
The initial
line identification in the \ion{Cu}{vi} spectrum was performed by 
\citet{poppeetal1974} 
which was considerably extended a few years later by 
\citet{raassenvankleef1981},
who used new exposures of a sliding spark discharge
to analyze the 
3d$^6$ - 3d$^5$4p and 
3d$^5$4s - 3d$^5$4p transition arrays. This
analysis led to the identification of all levels in the 
3d$^6$
configuration, except the highest 
$^1$S$_0$, 
as well as 
208 of the 214
levels in 
3d$^5$4p 
and 13 of the 74 levels in 
3d$^5$4s.  
Finally, the
analysis of the \ion{Cu}{vii} spectrum by
\citet{vanhethofetal1990} 
appeared too late to be included in the compilation of 
\citet{sugarmusgrove1990}. 
They determined all 37 levels of the 
3d$^5$ 
ground configuration
and 129 of the 180 possible levels of the 
3d$^4$4p 
configuration.  With
regard to the radiative parameters, very few studies have  focused
on the determination of electric dipole transition rates in \ion{Cu}{iv-vii}
ions so far. To our knowledge, the only available data were
recently obtained by 
\citet{aggarwaletal2016} 
who used the
quasi-relativistic approach (QR) with a very large configuration
interaction (CI) expansion to compute oscillator strengths and
transition probabilities in \ion{Cu}{vi}.

\paragraph{Oscillator strength calculations for the \ion{Cu}{iv-vii} ions.}

The method adopted in our work to model the atomic structures and
compute the radiative parameters in the \ion{Cu}{iv-vii} ions was the
pseudo-relativistic Hartree-Fock (HFR) approach originally introduced
by 
\citet{cowan1981} 
and modified to take core-polarization effects into
account, giving rise to the so-called HFR+CPOL method 
\citep[see, e.g.,][]{quinetetal1999,quinetetal2002,quinet2017}.

For \ion{Cu}{iv}, configuration
interaction was considered among the configurations
3d$^8$, 
3d$^7$4s, 
3d$^7$5s,
3d$^7$4d, 
3d$^7$5d, 
3d$^6$4s$^2$, 
3d$^6$4p$^2$, 
3d$^6$4d$^2$, 
3d$^6$4s5s, and 
3d$^6$4s4d for the
even parity, and 
3d$^7$4p, 
3d$^7$5p, 
3d$^7$4f, 
3d$^7$5f, 
3d$^6$4s4p, 
3d$^6$4s5p,
3d$^6$4s4f, and 
3d$^6$4p4d 
for the odd parity. The core-polarization
parameters were the dipole polarizability of a \ion{Cu}{vi} ionic core, as
reported by 
\citet{fragaetal1976}, 
i.e., $\alpha_\mathrm{d}$ = 0.67\,a.u\@., and the cut-off
radius, $r_c$ = 0.80 a.u, corresponding to the HFR mean value $<$$r$$>$ of the
outermost core orbital (3d). Using the experimental energy levels
compiled at NIST 
\citep{kramidaetal2019}, 
some radial integrals (average
energy, Slater, spin-orbit, and effective interaction parameters) of
3d$^8$, 
3d$^7$4s, 
3d$^7$5s, 
3d$^7$4d, 
3d$^6$4s$^2$, and 
3d$^7$4p 
configurations were
optimized by a well-established least-squares fitting procedure in
which the mean deviations were found to be equal to 206\,cm$^{-1}$ for the
even parity and 183\,cm$^{-1}$ for the odd parity. It is worth noting that
when performing the fit we found that the experimental energy level
at 306941.8\,cm$^{-1}$ was incorrectly classified as J = 1 in the NIST tables,
while according to our calculations this level should be designated
as J = 2.  

For \ion{Cu}{v}, the configurations included in the HFR model
were 
3d$^7$, 
3d$^6$4s, 
3d$^6$5s, 
3d$^6$4d, 
3d$^5$4s$^2$, 
3d$^5$4p$^2$, 
3d$^5$4d$^2$, and 
3d$^5$4s4d 
for
the even parity, and 
3d$^6$4p, 
3d$^6$5p, 
3d$^6$4f, 
3d$^5$4s4p, 
3d$^5$4s5p, and
3d$^5$4s4f 
for the odd parity. In this ion the semi-empirical process
was carried out to optimize the radial integrals corresponding to 
3d$^7$,
3d$^6$4s, and 
3d$^6$4p 
configurations using the experimental levels reported
in the NIST database 
\citep{kramidaetal2019}.
The mean deviations
between calculated and experimental energy levels were 325\,cm$^{-1}$ and
259\,cm$^{-1}$ for even and odd parities, respectively. Core-polarization
effects were estimated using a dipole polarizability and a cut-off
radius corresponding to a \ion{Cu}{vii} ionic core, i.e.,
$\alpha_\mathrm{d}$ = 0.47\,a.u\@. 
\citep{fragaetal1976}, 
and $r_c$ = 0.75\,a.u\@.  

In the case of \ion{Cu}{vi}, the
configuration interaction was considered among the following
configurations: 
3d$^6$, 
3d$^5$4s, 
3d$^5$5s, 
3d$^5$4d, 
3d$^5$5d, 
3d$^4$4s$^2$, 
3d$^4$4p$^2$,
3d$^4$4d$^2$, 
3d$^4$4s4d, 
3d$^4$4s5d (even parity) and 
3d$^5$4p, 
3d$^5$5p, 
3d$^5$4f,
3d$^4$4s4p, 
3d$^4$4s5p, 
3d$^4$4s4f 
(odd parity). The core-polarization
parameters were those corresponding to a \ion{Cu}{viii} ionic core, i.e.,
$\alpha_\mathrm{d}$ = 0.40\,a.u\@. 
\citep{fragaetal1976}, 
and $r_c$ = 0.72\,a.u\@. The radial
parameters of the
3d$^6$, 
3d$^5$4s, and 
3d$^5$4p 
configurations were optimized to
minimize the differences between the computed Hamiltonian eigenvalues
and the experimental energy levels listed at NIST 
\citep{kramidaetal2019} giving rise to mean deviations of 442\,cm$^{-1}$ (even parity) and
429\,cm$^{-1}$ (odd parity).  

Finally, for \ion{Cu}{vii}, ten even- and eight
odd-parity configurations were included in the HFR model to compute
the radiative parameters, i.e., 
3d$^5$, 
3d$^4$4s, 
3d$^4$5s, 
3d$^4$4d, 
3d$^4$5d,
3d$^3$4s$^2$, 
3d$^3$4p$^2$, 
3d$^3$4d$^2$, 
3d$^3$4s5s, 
3d$^3$4s4d, and 
3d$^4$4p, 
3d$^4$5p, 
3d$^4$4f,
3d$^4$5f, 
3d$^3$4s4p, 
3d$^3$4s5p, 
3d$^3$4s4f, and 
3d$^3$4p4d, 
respectively. An ionic core
of the type \ion{Cu}{ix} was considered to estimate the core-polarization
effects with the parameters 
$\alpha_\mathrm{d}$ = 0.34\,a.u\@. 
\citep{fragaetal1976} 
and $r_c$ = 0.70\,a.u\@. The semi-empirical optimization process was carried out to
adjust the radial parameters in 
3d$^5$ and 
3d$^4$4p 
with the experimental
energy levels taken from 
\citet{vanhethofetal1990} 
leading to average
energy deviations of 78\,cm$^{-1}$ and 365\,cm$^{-1}$ for even and odd parities,
respectively.  

The parameters that we had adopted for our computations are given in 
Tables \ref{tab:cuiv:para} - \ref{tab:cuvii:para} and a comparison of
calculated and experimental energies is shown in
Tables \ref{tab:cuiv:ener} - \ref{tab:cuvii:ener} for \ion{Cu}{iv-vii}, respectively.
In Tables \ref{tab:cuiv:loggf} - \ref{tab:cuvii:loggf}
(provided via the registered GAVO T\"ubingen Oscillator Strengths Service TOSS), 
we give the newly computed weighted
oscillator strengths ($\log gf$) and transition probabilities ($gA$, in
s$^{-1}$) together with the experimental values (in cm$^{-1}$) of the lower and
upper energy levels and the corresponding Ritz wavelengths (in \AA). In
the final column of each table we also give the cancellation factor
(CF), as defined by  \citet{cowan1981}. 

Let us remind that very low values of the CF 
(typically $<$ 0.05) indicate strong cancellation effects in
the calculation of line strengths. In these cases, the corresponding
$\log gf$ and $gA$ values could be very inaccurate and therefore need to
be considered with some care.  As mentioned above, no radiative rates
were previously published for the copper ions considered in our work
 except \ion{Cu}{vi}, for which theoretical oscillator strengths were
recently reported by 
\citet{aggarwaletal2016}, 
who used the
quasi-relativistic approach (QR) with a very large configuration
interaction expansion. When comparing these latter results with ours,
we found an overall good agreement, in particular for the strongest
lines with $\log gf > -1$, for which the mean deviation between the two sets
of data was found to be about 20\,\%, with a general tendency such that
our $\log gf$ values appear systematically slightly higher than those of
\citet{aggarwaletal2016}.

\section{Results and conclusions}
\label{sect:results}

We calculated oscillator strengths of \ion{Cu}{iv\,-\,vii} and compiled
oscillator strengths of 
\ion{In}{iv} \citep{swapniltauheed2013,ryabtsevkononov2016},
\ion{In}{v}  \citep{varshneytauheed2016,swapniltauheed2016,ryabtsev2018}, and
\ion{In}{vi} \citep{kononovetal2017,ryabtsevetal2018}
from the literature.
 These elements were included in our model-atmosphere calculations.

\begin{figure*}
   \resizebox{\hsize}{!}{\includegraphics{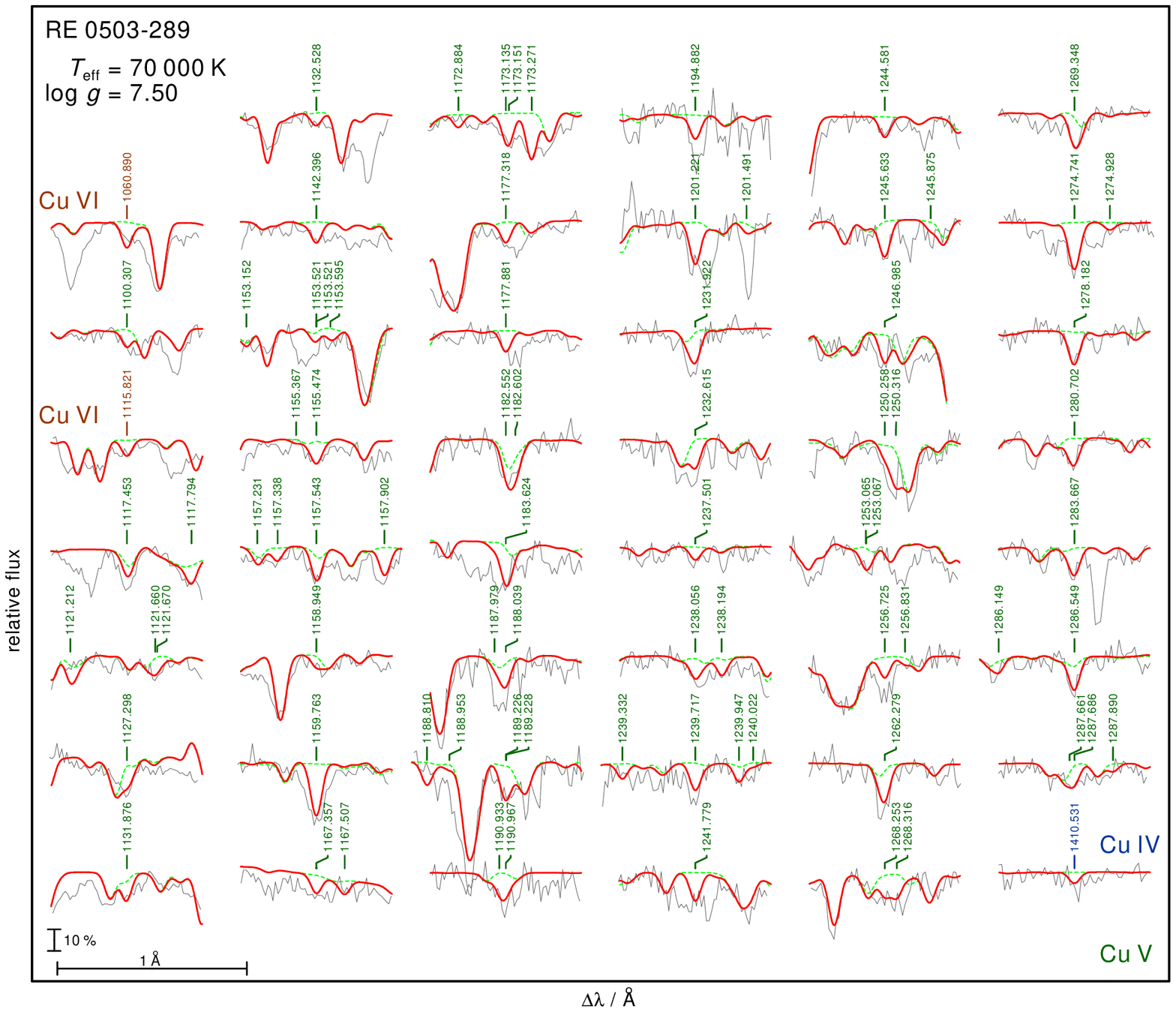}}
    \caption{Prominent Cu lines in the observation (gray line) of \re, 
             labeled with their wavelengths from Tables\,\ref{tab:cuiv:loggf}, \ref{tab:cuv:loggf}, and \ref{tab:cuvi:loggf}.
             \Jonww{Cu}{vi}{1060.890, 1100.307}, and \Jonw{Cu}{iv}{1410.531} are indicated with an 
             additional ion label,
             all other lines stem from \ion{Cu}{v}.
             For the identification of other lines that are visible in the spectrum, please visit
             \url{http://astro.uni-tuebingen.de/~TVIS/objects/RE0503-289}, our
             T\"ubingen VISualization Tool (TVIS).
             The thick red spectrum is calculated from our best model with a Cu mass fraction
             of $9.3 \times 10^{-5}$.
             The dashed green line shows a synthetic spectrum calculated without Cu. 
             The vertical bar indicates 10\,\% of the continuum flux.
            }
   \label{fig:cu_re}
\end{figure*}

\begin{figure*}
   \resizebox{\hsize}{!}{\includegraphics{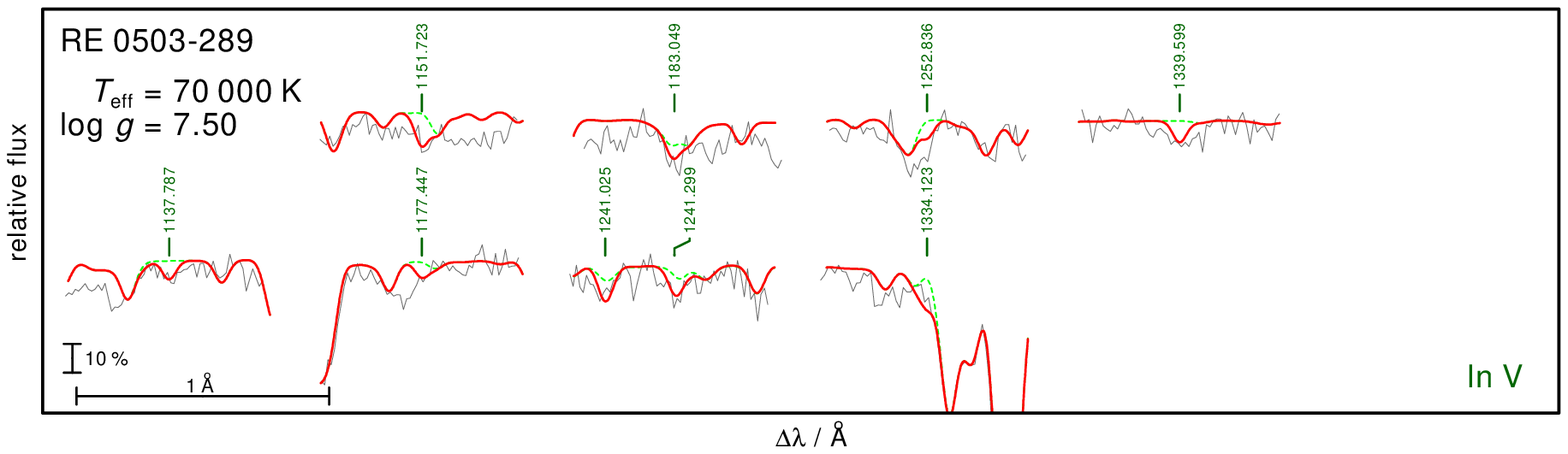}}
    \caption{Same as Fig.\,\ref{fig:cu_re}, but for In.
      Lines are labeled with their wavelengths from Table\,\ref{tab:inlineids_re}.
      The synthetic spectrum is calculated with an In mass fraction of $3.0 \times 10^{-6}$.
            }
   \label{fig:in_re}
\end{figure*}

\begin{figure*}
   \resizebox{\hsize}{!}{\includegraphics{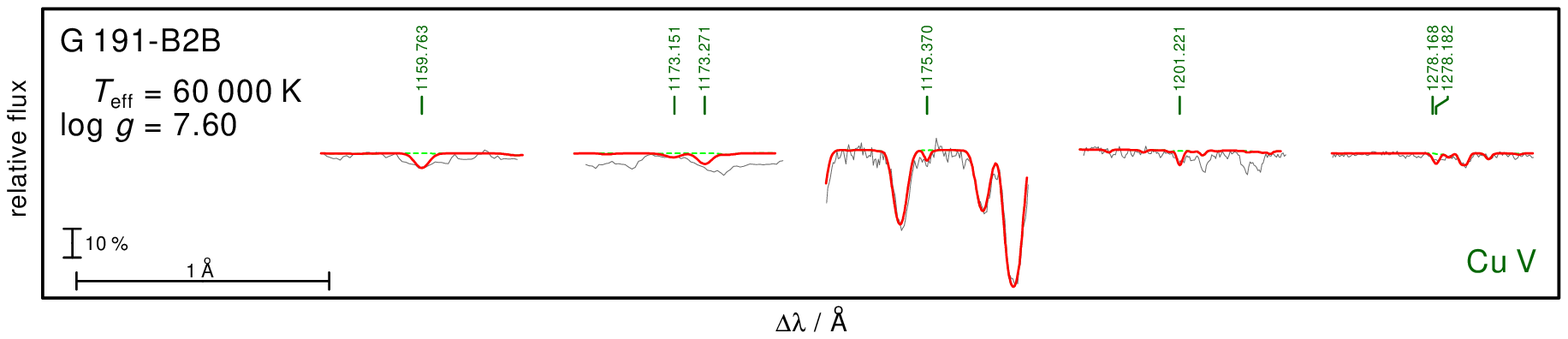}}
    \caption{Same as Fig.\,\ref{fig:cu_re}, but for \gb.
            }
   \label{fig:cu_gb}
\end{figure*}

To unambiguously identify lines of one individual species, we calculated two spectra
for each star (\re and \gb),
one with all opacities of that element included and one with its line opacities
switched off artificially (cf. Figs\@.\,\ref{fig:cu_re}, \ref{fig:in_re}, and \ref{fig:cu_gb}).
This enables us to identify Cu and In lines, even if they are blended with other lines  (cf.
Tables\,\ref{tab:culineids_re}, \ref{tab:inlineids_re}, and \ref{tab:cuinlineids_gb}) or
if they reduce the flux without exhibiting a complete line profile.
\vspace{2mm}

\noindent {\sc\bf \re.}~~We identified 
54 Cu lines 
(1 \ion{Cu}{iv}, 
51 \ion{Cu}{v}, 
and 2 \ion{Cu}{vi};
Fig.\,\ref{fig:cu_re}, Table\,\ref{tab:culineids_re})
and 5 \ion{In}{v} lines 
(Fig.\,\ref{fig:in_re}, Table\,\ref{tab:inlineids_re}).
From a detailed line-profile comparison of our model (\Teffw{70\,000} and \loggw{7.5}) 
with the UV observations, we found the best simultaneous agreement for all identified Cu and In lines at
abundances of 
$9.3^{+3.0}_{-2.0} \times 10^{-5}$ (mass fraction, 132 times solar) and
$3.0^{+0.5}_{-0.5} \times 10^{-6}$ (56\,600
times solar), respectively.
The \Teff and \logg error propagation is evaluated from two models at the error limits
with highest and lowest degree of ionization, i.e.,
\Teffw{72\,000} and \loggw{7.4}) and
\Teffw{68\,000} and \loggw{7.6}), respectively.
We found that this abundance error is smaller than 0.1\,dex. 
Finally, we adopted the above Cu and In mass fractions with uncertainties of 0.2\,dex.
\vspace{2mm}

\noindent {\sc\bf \gb.}~~The four strongest \ion{Cu}{v} lines and/or blends in the synthetic spectrum are identified 
in the observation (Fig.\,\ref{fig:cu_gb}, Table\,\ref{tab:cuinlineids_gb}). They are well reproduced 
by a model (\Teffw{60\,000}, \loggw{7.6}) with a Cu mass fraction of $6.3^{+3.0}_{-2.0} \times 10^{-6}$ (nine times solar).
The error estimation is performed analogously to that of \re (see above) and we found the same
uncertainty of 0.2\,dex.

\begin{figure}
   \resizebox{\columnwidth}{!}{\includegraphics{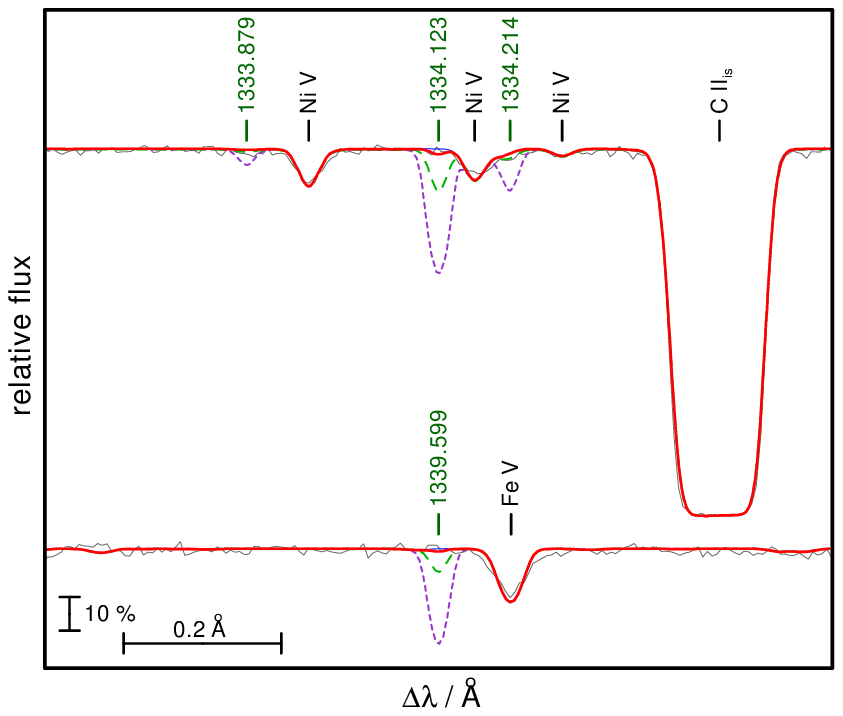}}
    \caption{STIS observation of \gb (gray) compared with synthetic line profiles of
             \ionw{In}{v}{1334.123} and
             \ionw{In}{vi}{1339.599} calculated with four In abundances:
             without (thin blue line),
             with 1000 times (thick red), 
                 10\,000 times (short-dashed violet), and
                100\,000 times solar abundance (long-dashed green).
            }
   \label{fig:ingb}
\end{figure}

No In line can be identified in the UV observation of \gb. We used two of the strongest lines,
namely \Ionww{In}{5}{1334.123, 1339.599}, to determine an upper abundance limit of
$5.3 \times 10^{-7}$ (mass fraction, 1000 times solar; Fig.\,\ref{fig:ingb}).
\vspace{2mm}

\begin{figure}
   \resizebox{\hsize}{!}{\includegraphics{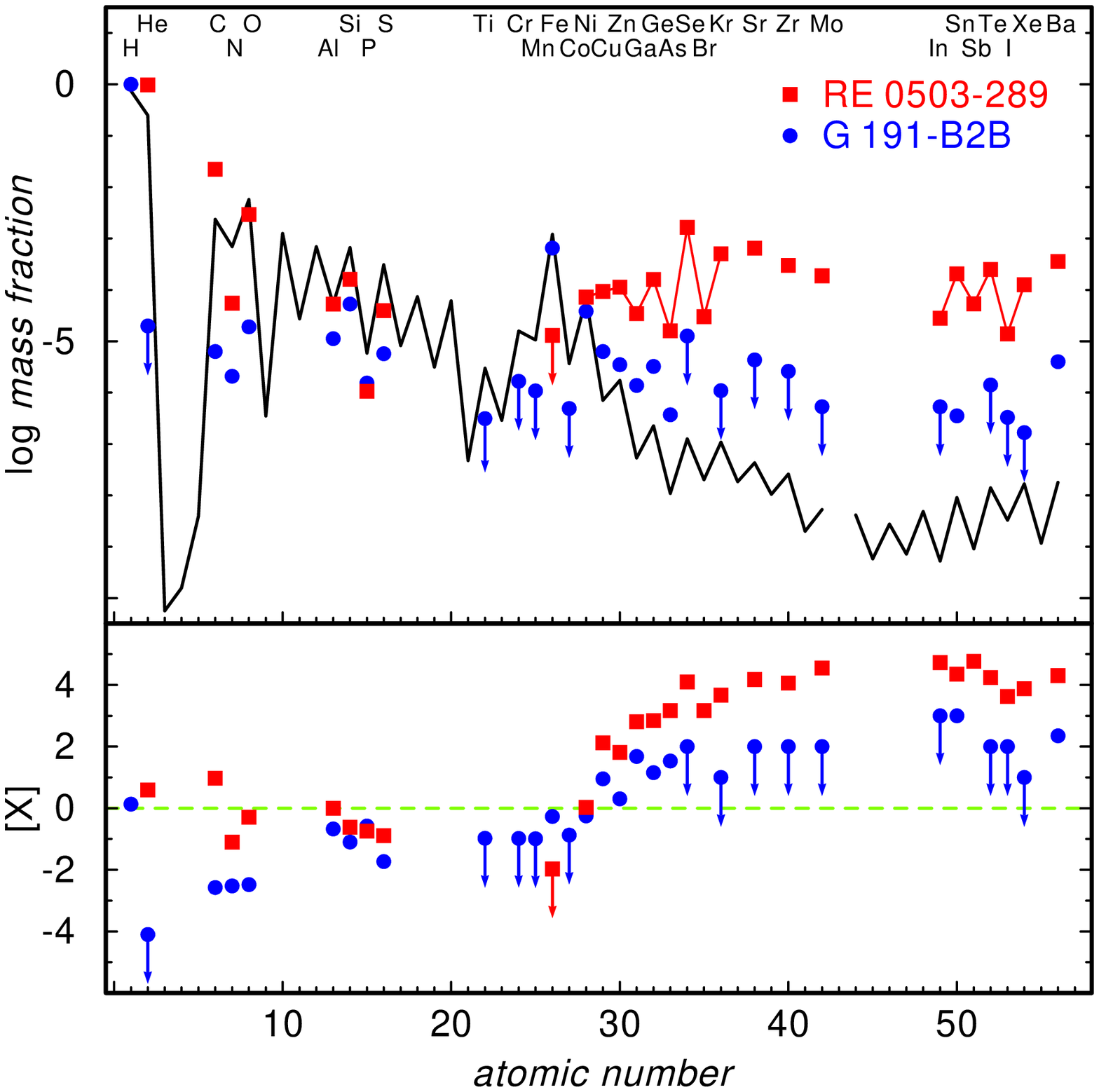}}
    \caption{Solar abundances \citep[thick black line]{asplundetal2009,scottetal2015a,scottetal2015b,grevesseetal2015}
             compared with the determined photospheric abundances of 
             \gb \citep[blue circles,][and this work]{rauchetal2013} and
             \re \citep[red squares,][and this work]{dreizlerwerner1996, rauchetal2012ge, rauchetal2014zn, rauchetal2014ba, rauchetal2015ga, rauchetal2016mo, rauchetal2016kr, rauchetal2016zr, rauchetal2017sesrtei}.
             The uncertainties of the abundances are about 0.2\,dex in general. Arrows indicate upper limits.
             Top panel: Abundances given as logarithmic mass fractions. 
                        Determined TIE abundances of subsequent species are combined with lines.
             Bottom panel: Abundance ratios to respective solar values, 
                           [X] denotes log (fraction\,/\,solar fraction) of species X.
                           The dashed green line indicates solar abundances.
            }
   \label{fig:X}
\end{figure}

To summarize, the determined Cu and In abundances closely match  the already known TIE abundance patterns in \gb and \re
(Fig.\ref{fig:X}), which is the result of effective radiative levitation \citep{rauchetal2016mo}. 
The TIE enrichment is much stronger in \re compared to that in \gb due to the lower \Teff of the latter
\citep[Fig.\,\ref{fig:evol}, \Teffw{60\,000 \pm 2000} and $\log g = 7.6 \pm 0.05$,][]{rauchetal2013}.
The relative TIE abundances in \re, however, obviously resemble the relative solar abundance ratios
(Fig.\ref{fig:X}, cf. $29 \le Z \le 36$ and $49 \le Z \le 54$), i.e., higher abundances for species 
with even $Z$ compared to those with   odd $Z$ \citep[Oddo-Harkins rule,][]{oddo1914,harkins1917}. 

Surface abundance patterns that are the result of diffusion (cf. Fig.\,\ref{fig:X}) and exhibit strong TIE
enrichment are also found recently in He-sdOs, e.g.,
\lsiv \citep{naslimetal2011}, in
\feige \citep{latouretal2019}, in
\hd and \hz \citep{dorschetal2019},
and in the DAO-type star \bd \citep{loeblingetal2019}.
However, \citet{hoyeretal2017} and \citet{rauchetal2019} have already mentioned that strong radiative 
levitation of trans-iron TIEs wipes out all information about their AGB abundances, and thus previous
stellar evolution. This is general for all species in all stars with a diffusion impact on their surface 
abundances.

Among the WDs, \re exhibits lines from an unrivaled number of TIEs:  18 species
(\url{http://astro.uni-tuebingen.de/~TVIS/objects/RE0503-289}). Many more lines in the UV
spectral region remain unidentified. Calculations of transition probabilities of other TIEs are necessary to make
further progress.

\begin{acknowledgements}
The GAVO project had been supported by the Federal Ministry of Education and Research (BMBF) 
at T\"ubingen (05\,AC\,6\,VTB, 05\,AC\,11\,VTB) and is funded
at Heidelberg (05\,A\,17\,VH2).
Financial support from the Belgian FRS-FNRS is also acknowledged. 
PQ is research director of this organization.
Some of the data presented in this paper were obtained from the
Mikulski Archive for Space Telescopes (MAST). STScI is operated by the
Association of Universities for Research in Astronomy, Inc., under NASA
contract NAS5-26555. Support for MAST for non-HST data is provided by
the NASA Office of Space Science via grant NNX09AF08G and by other
grants and contracts. 
The TIRO (\url{http://astro.uni-tuebingen.de/~TIRO}),
    TMAD (\url{http://astro.uni-tuebingen.de/~TMAD}), 
    TOSS (\url{http://astro.uni-tuebingen.de/~TOSS}), and
    TVIS (\url{http://astro.uni-tuebingen.de/~TVIS}) tools and services
were constructed as part of the T\"ubingen project 
(https://uni-tuebingen.de/de/122430)
of the German Astrophysical Virtual Observatory
(GAVO, \url{http://www.g-vo.org}).
This research has made use of 
NASA's Astrophysics Data System and
the SIMBAD database, operated at CDS, Strasbourg, France.
\end{acknowledgements}

\bibliographystyle{aa}
\bibliography{36620}

\begin{appendix}

\onecolumn

\section{Additional tables}
\label{app:addtabs}

\setlength{\tabcolsep}{.2em}
\begin{longtable}{r@{\,}lr@{.}lr@{.}lcccrcccp{5.4cm}}
\caption{\label{tab:culineids_re}Identified \ion{Cu}{iv-vi} lines in the UV spectrum of \re.} \\
\hline\hline
\noalign{\smallskip}
&&&&&                 & \multicolumn{3}{c}{Lower level} && \multicolumn{3}{c}{Upper level} &  \\
\cline{7-9}\cline{11-13}
\noalign{\smallskip}
\multicolumn{2}{c}{Ion} & 
\multicolumn{2}{c}{Wavelength} & 
\multicolumn{2}{c}{$\log g_\mathrm{i}f_\mathrm{ik}$} &
Energy & 
Parity & 
$J$  && 
Energy & 
Parity & $J$ & 
Comment \\
&& 
\multicolumn{2}{c}{\AA} & 
\multicolumn{2}{c}{~} &
cm$^{-1}$ & 
& 
&& 
cm$^{-1}$ & 
& & \\
\noalign{\smallskip}
\hline                                                                                            
\endfirsthead
\caption{continued.}\\
\hline\hline
\noalign{\smallskip}
&&&&&                 & \multicolumn{3}{c}{Lower level} && \multicolumn{3}{c}{Upper level} &  \\
\cline{7-9}\cline{11-13}
\noalign{\smallskip}
\multicolumn{2}{c}{Ion} & 
\multicolumn{2}{c}{Wavelength} & 
\multicolumn{2}{c}{$\log g_\mathrm{i}f_\mathrm{ik}$} &
Energy & 
Parity & 
$J$  && 
Energy & 
Parity & $J$ & 
Comment \\
&& 
\multicolumn{2}{c}{\AA} & 
\multicolumn{2}{c}{~} &
cm$^{-1}$ & 
& 
&& 
cm$^{-1}$ & 
& & \\
\noalign{\smallskip}
\hline
\noalign{\smallskip}
\endhead
\hline
\noalign{\smallskip}
\endfoot
\noalign{\smallskip}
\hline                                                                                            
\noalign{\smallskip}
Cu&{\sc vi} & \hbox{}\hspace{5mm}   1060& 890 &     0& 38 &  250877 &  e &  3.0 & &  345137 &  o &  4.0 & \\
Cu&{\sc v}  &                       1100& 307 &  $-$0& 52 &  187779 &  e &  4.5 & &  278663 &  o &  3.5 & \\
Cu&{\sc vi} &                       1115& 821 &     0& 24 &  265639 &  e &  2.0 & &  355259 &  o &  3.0 & \\
Cu&{\sc v}  &                       1117& 453 &  $-$0& 50 &  190100 &  e &  1.5 & &  279590 &  o &  1.5 & blend with \ionw{Zn}{v}{1117.466} \\
  &         &                       1117& 794 &  $-$0& 28 &  188833 &  e &  3.5 & &  278295 &  o &  2.5 & \\
  &         &                       1121& 212 &  $-$0& 39 &  190400 &  e &  0.5 & &  279590 &  o &  1.5 & too strong in model \\
  &         &                       1121& 660 &  $-$0& 06 &  228048 &  e &  4.5 & &  317201 &  o &  4.5 & blend with \ionw{Cu}{v}{1121.670} \\
  &         &                       1121& 670 &  $-$0& 22 &  265752 &  e &  4.5 & &  354905 &  o &  3.5 & blend with \ionw{Cu}{v}{1121.660} \\
\hline
\end{longtable}

\twocolumn

\begin{table*}\centering
\setlength{\tabcolsep}{0.3em}
\caption{Same as Table\,\ref{tab:culineids_re}, but for \ion{In}{v}.}
\label{tab:inlineids_re}
\begin{tabular}{rlr@{.}lr@{.}lcccrcccp{5.4cm}}
\hline\hline
&&&&&                 & \multicolumn{3}{c}{Lower level} && \multicolumn{3}{c}{Upper level} &                  \\
\cline{7-9}\cline{11-13}
\noalign{\smallskip}
\multicolumn{2}{c}{Ion} & 
\multicolumn{2}{c}{Wavelength / \AA} & 
\multicolumn{2}{c}{$\log g_\mathrm{i}f_\mathrm{ik}$} &
Energy\,/\,cm$^{-1}$ & 
Parity & 
$J$  && 
Energy\,/\,cm$^{-1}$ & 
Parity & $J$  & 
Comment \\
\noalign{\smallskip}
\hline                                                                                            
\noalign{\smallskip}
In&{\sc v}  & \hbox{}\hspace{5mm}   1137&      787 &$-$0& 09 & 163980 & e & 3.5 & & 251868 & o & 4.5 & \\
  &         &                       1151&      723 &   0& 33 & 160428 & e & 4.5 & & 247254 & o & 4.5 & blend with \ionw{Zn}{v}{1151.787} \\
  &         &                       1177&      447 &$-$0& 01 & 167761 & e & 2.5 & & 252691 & o & 2.5 & weak \\
  &         &                       1183&      049 &   0& 11 & 163980 & e & 3.5 & & 248507 & o & 3.5 & blend with \newline \ionww{Zn}{v}{1183.041, 1183.100} \newline
                                                                                                                            \ionw{Ga}{v}{1183.110} \\
  &         &                       1241&      025 &   0& 39 & 171290 & e & 3.5 & & 251868 & o & 4.5 & blend with \newline \ionww{Ni}{v}{1241.015, 1241.047} \\
  &         &                       1241&      299 &   0& 59 & 188396 & e & 4.5 & & 268957 & o & 5.5 & blend with \newline \ionww{O}{iii}{1241.125, 1241.176} \\
  &         &                       1252&      836 &   0& 03 & 163980 & e & 3.5 & & 243798 & o & 3.5 & blend with \ionw{Ni}{v}{1252.765} \\
  &         &                       1334&      123 &   0& 37 & 160428 & e & 4.5 & & 235384 & o & 3.5 & blend with interstellar \ion{C}{ii}, weak \\
  &         &                       1339&      599 &   0& 17 & 163980 & e & 3.5 & & 238629 & o & 4.5 & \\
\hline
\end{tabular}
\tablefoot{
  The wavelengths, weighted oscillator strengths, and energies correspond to those in
  \citet{varshneytauheed2016,swapniltauheed2016} and \citet{ryabtsev2018}.
}
\end{table*}

\begin{table*}\centering
\setlength{\tabcolsep}{0.3em}
\caption{Same as Table\,\ref{tab:culineids_re}, but for \gb.}
\label{tab:cuinlineids_gb}
\begin{tabular}{rlr@{.}lr@{.}lcccrcccp{5.4cm}}
\hline\hline
&&&&&                 & \multicolumn{3}{c}{Lower level} && \multicolumn{3}{c}{Upper level} &                  \\
\cline{7-9}\cline{11-13}
\noalign{\smallskip}
\multicolumn{2}{c}{Ion} & 
\multicolumn{2}{c}{Wavelength / \AA} & 
\multicolumn{2}{c}{$\log g_\mathrm{i}f_\mathrm{ik}$} &
Energy\,/\,cm$^{-1}$ & 
Parity & 
$J$  && 
Energy\,/\,cm$^{-1}$ & 
Parity & $J$  & 
Comment \\
\noalign{\smallskip}
\hline                                                                                            
\noalign{\smallskip}
Cu&{\sc v} & \hbox{}\hspace{5mm}1159&763 &    0& 52 & 187779 & e & 4.5 & & 274004 & o & 5.5 & \\
  &        &                    1173&151 & $-$0& 22 & 188833 & e & 3.5 & & 274073 & o & 3.5 & weak \\
  &        &                    1173&271 &    0& 39 & 188833 & e & 3.5 & & 274065 & o & 4.5 & blend with unidentified line \\
  &        &                    1175&370 &    0& 09 & 199441 & e & 3.5 & & 284521 & o & 2.5 & \\
  &        &                    1201&221 &    0& 64 & 220208 & e & 6.5 & & 303456 & o & 7.5 & \\
  &        &                    1278&168 & $-$0& 49 & 229588 & e & 3.5 & & 307825 & o & 4.5 & blend with \ionw{Cu}{v}{1278.182} \\
  &        &                    1278&182 & $-$0& 04 & 188833 & e & 3.5 & & 267069 & o & 2.5 & blend with \ionw{Cu}{v}{1278.168} \\
\hline
\end{tabular}
\end{table*}

\begin{table}\centering
\setlength{\tabcolsep}{0.3em}
\caption{Ions with recently calculated oscillator strengths.}
\label{tab:tee}
\begin{tabular}{rrll}
\hline
\hline
Cu & {\sc iv} &-\hspace{2mm}{\sc vii}          & Rauch et al. {this paper} \\
Zn & {\sc iv} &-\hspace{2mm}{\sc v}            & \citet{rauchetal2014zn} \\
Ga & {\sc iv} &-\hspace{2mm}{\sc vi}           & \citet{rauchetal2015ga} \\
Ge & {\sc  v} &-\hspace{2mm}{\sc vi}           & \citet{rauchetal2012ge} \\
Se & {\sc  v} &                                & \citet{rauchetal2017sesrtei} \\
Kr & {\sc iv} &-\hspace{2mm}{\sc vii}          & \citet{rauchetal2016kr} \\
Sr & {\sc iv} &-\hspace{2mm}{\sc vii}          & \citet{rauchetal2017sesrtei} \\
Zr & {\sc iv} &-\hspace{2mm}{\sc vii}          & \citet{rauchetal2016zr} \\
Mo & {\sc iv} &-\hspace{2mm}{\sc vii}          & \citet{rauchetal2016mo} \\
Te & {\sc vi} &                                & \citet{rauchetal2017sesrtei} \\
I  & {\sc vi} &                                &\citet{rauchetal2017sesrtei} \\
Xe & {\sc iv} &-\hspace{2mm}{\sc v}, {\sc vii} & \citet{rauchetal2015xe,rauchetal2016zr} \\
Ba & {\sc  v} &-\hspace{2mm}{\sc vii}          & \citet{rauchetal2014ba} \\
\hline
\end{tabular}
\end{table}

\clearpage
\onecolumn

\begin{longtable}{llrrrl}
\caption{\label{tab:cuiv:para}Radial parameters (in cm$^{-1}$) adopted for the calculations in \ion{Cu}{iv}.} \\
\hline\hline
\noalign{\smallskip}
Configuration & Parameter & HFR & Fitted & Ratio & Note$^a$ \\ 
\noalign{\smallskip}
\hline
\endfirsthead
\caption{continued.}\\
\hline\hline
\noalign{\smallskip}
Configuration & Parameter & HFR & Fitted & Ratio & Note$^a$ \\
\noalign{\smallskip}
\hline
\noalign{\smallskip}
\endhead
\hline
\noalign{\smallskip}
\endfoot
\noalign{\smallskip}
\multicolumn{6}{c}{Even parity} \\
3d$^8$           & E$_{av}$          &   17190 &  16729 &       &    \\
                 & F$^2$(3d,3d)      &  112092 &  98330 & 0.877 &    \\ 
                 & F$^4$(3d,3d)      &   69989 &  60209 & 0.860 &    \\
                 & $\alpha$          &       0 &     91 &       &    \\
                 & $\beta$           &       0 &   -237 &       &    \\
                 & $\zeta$$_{3d}$    &     903 &    910 & 1.007 &    \\
3d$^7$4s         & E$_{av}$          &  147324 & 152124 &       &    \\
\hline
\end{longtable}
$^a$ F: parameter fixed to its {\it ab initio} value; R: ratios of these parameters have been fixed in the fitting process\\

\begin{longtable}{llrrrl}
\caption{\label{tab:cuv:para}Radial parameters (in cm$^{-1}$) adopted for the calculations in \ion{Cu}{v}.} \\
\hline\hline
\noalign{\smallskip}
Configuration & Parameter & HFR & Fitted & Ratio & Note$^a$ \\ 
\noalign{\smallskip}
\hline
\endfirsthead
\caption{continued.}\\
\hline\hline
\noalign{\smallskip}
Configuration & Parameter & HFR & Fitted & Ratio & Note$^a$ \\
\noalign{\smallskip}
\hline
\noalign{\smallskip}
\endhead
\hline
\noalign{\smallskip}
\endfoot
\noalign{\smallskip}
\multicolumn{6}{c}{Even parity} \\
3d$^7$           & E$_{av}$          &   30899 &  30322 &       &    \\
                 & F$^2$(3d,3d)      &  119816 & 105551 & 0.881 &    \\ 
                 & F$^4$(3d,3d)      &   75176 &  65503 & 0.871 &    \\
                 & $\alpha$          &       0 &    109 &       &    \\
                 & $\beta$           &       0 &   -587 &       &    \\
                 & $\zeta$$_{3d}$    &     989 &   1002 & 1.013 &    \\
3d$^6$4s         & E$_{av}$          &  234957 & 237924 &       &    \\
\hline
\end{longtable}

\begin{longtable}{llrrrl}
\caption{\label{tab:cuvi:para}Radial parameters (in cm$^{-1}$) adopted for the calculations in \ion{Cu}{vi}.} \\
\hline\hline
\noalign{\smallskip}
Configuration & Parameter & HFR & Fitted & Ratio & Note$^a$ \\ 
\noalign{\smallskip}
\hline
\endfirsthead
\caption{continued.}\\
\hline\hline
\noalign{\smallskip}
Configuration & Parameter & HFR & Fitted & Ratio & Note$^a$ \\
\noalign{\smallskip}
\hline
\noalign{\smallskip}
\endhead
\hline
\noalign{\smallskip}
\endfoot
\noalign{\smallskip}
\multicolumn{6}{c}{Even parity} \\
3d$^6$           & E$_{av}$          &   47058 &  46817 &       &    \\
                 & F$^2$(3d,3d)      &  127114 & 112843 & 0.888 &    \\ 
                 & F$^4$(3d,3d)      &   80096 &  69557 & 0.868 &    \\
                 & $\alpha$          &       0 &    103 &       &    \\
                 & $\beta$           &       0 &   -264 &       &    \\
                 & $\zeta$$_{3d}$    &    1080 &   1142 & 1.057 &    \\
3d$^5$4s         & E$_{av}$          &  332771 & 334360 &       &    \\
\hline
\end{longtable}
$^a$ F: parameter fixed to its {\it ab initio} value\\

\begin{longtable}{llrrrl}
\caption{\label{tab:cuvii:para}Radial parameters (in cm$^{-1}$) adopted for the calculations in \ion{Cu}{vii}.} \\
\hline\hline
\noalign{\smallskip}
Configuration & Parameter & HFR & Fitted & Ratio & Note$^a$ \\ 
\noalign{\smallskip}
\hline
\endfirsthead
\caption{continued.}\\
\hline\hline
\noalign{\smallskip}
Configuration & Parameter & HFR & Fitted & Ratio & Note$^a$ \\
\noalign{\smallskip}
\hline
\noalign{\smallskip}
\endhead
\hline
\noalign{\smallskip}
\endfoot
\noalign{\smallskip}
\multicolumn{6}{c}{Even parity} \\
3d$^5$           & E$_{av}$          &   78756 &  78522 &       &    \\
                 & F$^2$(3d,3d)      &  134086 & 119456 & 0.891 &    \\ 
                 & F$^4$(3d,3d)      &   84807 &  74397 & 0.877 &    \\
                 & $\alpha$          &       0 &    118 &       &    \\
                 & $\beta$           &       0 &   -512 &       &    \\
                 & $\zeta$$_{3d}$    &    1176 &   1202 & 1.022 &    \\\\[0.3cm]
\multicolumn{6}{c}{Odd parity} \\
3d$^4$4p         & E$_{av}$          &  553653 & 554223 &       &    \\
                 & F$^2$(3d,3d)      &  139722 & 123787 & 0.886 &    \\ 
                 & F$^4$(3d,3d)      &   88617 &  78219 & 0.883 &    \\
                 & $\alpha$          &       0 &    123 &       &    \\
                 & $\beta$           &       0 &   -380 &       &    \\
                 & $\zeta$$_{3d}$    &    1263 &   1121 & 0.888 &    \\
                 & $\zeta$$_{4p}$    &    2481 &   2647 & 1.067 &    \\
                 & F$^2$(3d,4p)      &   35417 &  34983 & 0.988 &    \\
                 & G$^1$(3d,4p)      &   11725 &  11199 & 0.955 &    \\
                 & G$^3$(3d,4p)      &   11316 &  11458 & 1.013 &    \\ 
\hline
\end{longtable}

\begin{longtable}{rrrrl}
\caption{\label{tab:cuiv:ener} Comparison between available experimental and calculated energy levels in \ion{Cu}{iv}. Energies are given in cm$^{-1}$.} \\
\hline\hline
\noalign{\smallskip}
E$_{exp}$$^a$  & E$_{calc}$$^b$ & $\Delta$E & J & Leading components (in \%) in $LS$ coupling$^c$ \\ 
\noalign{\smallskip}
\hline
\endfirsthead
\caption{continued.}\\
\hline\hline
\noalign{\smallskip}
E$_{exp}$$^a$  & E$_{calc}$$^b$ & $\Delta$E & J & Leading components (in \%) in $LS$ coupling$^c$ \\
\noalign{\smallskip}
\hline
\noalign{\smallskip}
\endhead
\hline
\noalign{\smallskip}
\endfoot
\noalign{\smallskip}
\multicolumn{5}{c}{Even parity} \\
     0.0        &            8  &       $-$8    &4&     99      3d$^8$ $^3$F    \\
  1861.4        &         1857  &          4    &3&     99      3d$^8$ $^3$F    \\
  3077.6        &         3074  &          4    &2&     98      3d$^8$ $^3$F    \\
 16248.0        &        16238  &         10    &2&     82      3d$^8$ $^1$D    +       17      3d$^8$ $^3$P    \\
 19696.6        &        19750  &      $-$53    &2&     82      3d$^8$ $^3$P    +       17      3d$^8$ $^1$D    \\
 20096.6        &        20072  &         25    &1&     99      3d$^8$ $^3$P    \\
 20422.6        &        20405  &         18    &0&     99      3d$^8$ $^3$P    \\
\hline
\end{longtable}
\noindent
$^a$ From NIST compilation \citep{kramidaetal2019}.\\
$^b$ This work.\\
$^c$ Only the first three components higher than or equal to 5\% are given.

\begin{longtable}{rrrrl}
\caption{\label{tab:cuv:ener} Comparison between available experimental and calculated energy levels in \ion{Cu}{v}. Energies are given in cm$^{-1}$.} \\
\hline\hline
\noalign{\smallskip}
E$_{exp}$$^a$  & E$_{calc}$$^b$ & $\Delta$E & J & Leading components (in \%) in $LS$ coupling$^c$ \\ 
\noalign{\smallskip}
\hline
\endfirsthead
\caption{continued.}\\
\hline\hline
\noalign{\smallskip}
E$_{exp}$$^a$  & E$_{calc}$$^b$ & $\Delta$E & J & Leading components (in \%) in $LS$ coupling$^c$ \\
\noalign{\smallskip}
\hline
\noalign{\smallskip}
\endhead
\hline
\noalign{\smallskip}
\endfoot
\noalign{\smallskip}
\multicolumn{5}{c}{Even parity} \\
     0.0        &          136  &     $-$136    &4.5&   99      3d$^7$ $^4$F   \\
  1615.9        &         1715  &      $-$99    &3.5&   100     3d$^7$ $^4$F    \\
  2759.3        &         2840  &      $-$81    &2.5&   99      3d$^7$ $^4$F    \\
  3528.1        &         3596  &      $-$68    &1.5&   99      3d$^7$ $^4$F    \\
 20826.8        &        20622  &        205    &2.5&   99      3d$^7$ $^4$P    \\
 21065.9        &        20952  &        114    &1.5&   91      3d$^7$ $^4$P    +       8       3d$^7$ $^2$P    \\
 21935.1        &        21818  &        117    &0.5&   96      3d$^7$ $^4$P    \\
\hline
\end{longtable}
\noindent
$^a$ From NIST compilation \citep{kramidaetal2019}.\\
$^b$ This work.\\
$^c$ Only the first three components greater than or equal to 5\% are given.

\begin{longtable}{rrrrl}
\caption{\label{tab:cuvi:ener} Comparison between available experimental and calculated energy levels in \ion{Cu}{vi}. Energies are given in cm$^{-1}$.} \\
\hline\hline
\noalign{\smallskip}
E$_{exp}$$^a$  & E$_{calc}$$^b$ & $\Delta$E & J & Leading components (in \%) in $LS$ coupling$^c$ \\ 
\noalign{\smallskip}
\hline
\endfirsthead
\caption{continued.}\\
\hline\hline
\noalign{\smallskip}
E$_{exp}$$^a$  & E$_{calc}$$^b$ & $\Delta$E & J & Leading components (in \%) in $LS$ coupling$^c$ \\
\noalign{\smallskip}
\hline
\noalign{\smallskip}
\endhead
\hline
\noalign{\smallskip}
\endfoot
\noalign{\smallskip}
\multicolumn{5}{c}{Even parity} \\
     0.0        &        $-$84  &         84    &4&     99      3d$^6$ $^5$D    \\
  1195.8        &         1124  &         72    &3&     100     3d$^6$ $^5$D    \\
  1986.8        &         1925  &         62    &2&     99      3d$^6$ $^5$D    \\
  2489.4        &         2435  &         54    &1&     99      3d$^6$ $^5$D    \\
  2733.4        &         2683  &         50    &0&     99      3d$^6$ $^5$D    \\
 29285.0        &        29833  &     $-$548    &2&     60      3d$^6$ $^3$P    +       38      3d$^6$ $^3$P    \\
 30417.6        &        30593  &     $-$175    &6&     99      3d$^6$ $^3$H    \\
\hline
\end{longtable}
\noindent
$^a$ From NIST compilation \citep{kramidaetal2019}.\\
$^b$ This work.\\
$^c$ Only the first three components greater than or equal to 5\% are given.

\begin{longtable}{rrrrl}
\caption{\label{tab:cuvii:ener} Comparison between available experimental and calculated energy levels in \ion{Cu}{vii}. Energies are given in cm$^{-1}$.} \\
\hline\hline
\noalign{\smallskip}
E$_{exp}$$^a$  & E$_{calc}$$^b$ & $\Delta$E & J & Leading components (in \%) in $LS$ coupling$^c$ \\ 
\noalign{\smallskip}
\hline
\endfirsthead
\caption{continued.}\\
\hline\hline
\noalign{\smallskip}
E$_{exp}$$^a$  & E$_{calc}$$^b$ & $\Delta$E & J & Leading components (in \%) in $LS$ coupling$^c$ \\
\noalign{\smallskip}
\hline
\noalign{\smallskip}
\endhead
\hline
\noalign{\smallskip}
\endfoot
\noalign{\smallskip}
\multicolumn{5}{c}{Even parity} \\
     0.0        &        $-$13  &         13    &2.5&   100     3d$^5$ $^6$S    \\
 46442.7        &        46565  &     $-$122    &5.5&   100     3d$^5$ $^4$G    \\
 46515.3        &        46383  &        132    &2.5&   99      3d$^5$ $^4$G    \\
 46575.0        &        46604  &      $-$29    &4.5&   100     3d$^5$ $^4$G    \\
 46578.2        &        46520  &         58    &3.5&   99      3d$^5$ $^4$G    \\
 50731.7        &        50742  &      $-$10    &2.5&   86      3d$^5$ $^4$P    +       12      3d$^5$ $^4$D    \\
 51066.6        &        51059  &          8    &1.5&   89      3d$^5$ $^4$P    +       10      3d$^5$ $^4$D    \\
\hline
\end{longtable}
\noindent
$^a$ From \citet{ vanhethofetal1990}.\\
$^b$ This work.\\
$^c$ Only the first three components higher than or equal to 5\% are given.

\begin{longtable}{rrccrrccrrr}
\caption{\label{tab:cuiv:loggf}Calculated HFR oscillator strengths ($\log g_\mathrm{i}f_\mathrm{ik}$) and transition probabilities ($g_\mathrm{k}A_\mathrm{ki}$) 
                               in \ion{Cu}{iv}. CF is the
                               absolute value of the cancellation factor as defined by \citet{cowan1981}. 
                               In Cols\@. 3 and 6, e represents even and o odd.
                               This table is only available via the GAVO service TOSS (\url{http://dc.g-vo.org/TOSS}).} \\
\hline\hline
\noalign{\smallskip}
                 & \multicolumn{3}{c}{Lower level} && \multicolumn{3}{c}{Upper level} &           &      &    \\
\cline{2-4}\cline{6-8}
\noalign{\smallskip}
Wavelength{$^a$} / \AA & Energy{$^b$} / cm$^{-1}$ & Parity & $J$  && Energy{$^b$} / cm$^{-1}$ & Parity & $J$  & $\log g_\mathrm{i}f_\mathrm{ik}$ & $g_\mathrm{k}A_\mathrm{ki}$ / s$^{-1}$  & CF \\
\noalign{\smallskip}
\hline
\endfirsthead
\caption{continued.}\\
\hline\hline
\noalign{\smallskip}
                 & \multicolumn{3}{c}{Lower level} && \multicolumn{3}{c}{Upper level} &           &      &    \\
\cline{2-4}\cline{6-8}
\noalign{\smallskip}
Wavelength{$^a$} / \AA & Energy{$^b$} / cm$^{-1}$ & Parity & $J$  && Energy{$^b$} / cm$^{-1}$ & Parity & $J$  & $\log g_\mathrm{i}f_\mathrm{ik}$ & $g_\mathrm{k}A_\mathrm{ki}$ / s$^{-1}$  & CF \\
\noalign{\smallskip}
\hline
\noalign{\smallskip}
\endhead
\hline
\noalign{\smallskip}
\endfoot
\noalign{\smallskip}
\hline
\noalign{\smallskip}
\multicolumn{11}{p{15.4cm}}{{Notes.\,
                             $^{(a)}$\,All wavelengths (given in vacuum for                         $\lambda <   2000\,\mathrm{\AA}$, 
                                                                   air for  $2000\,\mathrm{\AA} \le \lambda \le 20000\,\mathrm{\AA}$,
                                                                vacuum for $20000\,\mathrm{\AA} <   \lambda$) are deduced from experimental energy levels.}
                             $^{(b)}$\,Experimental energy levels taken from \citet{kramidaetal2019}}.\\
\end{longtable}

\begin{longtable}{rrccrrccrrr}
\caption{\label{tab:cuv:loggf}Calculated HFR oscillator strengths ($\log g_\mathrm{i}f_\mathrm{ik}$) and transition probabilities ($g_\mathrm{k}A_\mathrm{ki}$) 
                               in \ion{Cu}{v}. CF is the
                               absolute value of the cancellation factor as defined by \citet{cowan1981}. 
                               In Cols\@. 3 and 6, e represents even and o odd.
                               This table is only available via the GAVO service TOSS (\url{http://dc.g-vo.org/TOSS}).} \\
\hline\hline
\noalign{\smallskip}
                 & \multicolumn{3}{c}{Lower level} && \multicolumn{3}{c}{Upper level} &           &      &    \\
\cline{2-4}\cline{6-8}
\noalign{\smallskip}
Wavelength{$^a$} / \AA & Energy{$^b$} / cm$^{-1}$ & Parity & $J$  && Energy{$^b$} / cm$^{-1}$ & Parity & $J$  & $\log g_\mathrm{i}f_\mathrm{ik}$ & $g_\mathrm{k}A_\mathrm{ki}$ / s$^{-1}$  & CF \\
\noalign{\smallskip}
\hline
\endfirsthead
\caption{continued.}\\
\hline\hline
\noalign{\smallskip}
                 & \multicolumn{3}{c}{Lower level} && \multicolumn{3}{c}{Upper level} &           &      &    \\
\cline{2-4}\cline{6-8}
\noalign{\smallskip}
Wavelength{$^a$} / \AA & Energy{$^b$} / cm$^{-1}$ & Parity & $J$  && Energy{$^b$} / cm$^{-1}$ & Parity & $J$  & $\log g_\mathrm{i}f_\mathrm{ik}$ & $g_\mathrm{k}A_\mathrm{ki}$ / s$^{-1}$  & CF \\
\noalign{\smallskip}
\hline
\noalign{\smallskip}
\endhead
\hline
\noalign{\smallskip}
\endfoot
\noalign{\smallskip}
\hline
\noalign{\smallskip}
\multicolumn{11}{p{15.4cm}}{{Notes.\,
                             $^{(a)}$\,All wavelengths (given in vacuum for                         $\lambda <   2000\,\mathrm{\AA}$, 
                                                                   air for  $2000\,\mathrm{\AA} \le \lambda \le 20000\,\mathrm{\AA}$,
                                                                vacuum for $20000\,\mathrm{\AA} <   \lambda$) are deduced from experimental energy levels.}
                             $^{(b)}$\,Experimental energy levels taken from \citet{kramidaetal2019}}.\\
\end{longtable}

\begin{longtable}{rrccrrccrrr}
\caption{\label{tab:cuvi:loggf}Calculated HFR oscillator strengths ($\log g_\mathrm{i}f_\mathrm{ik}$) and transition probabilities ($g_\mathrm{k}A_\mathrm{ki}$) 
                               in \ion{Cu}{vi}. CF is the
                               absolute value of the cancellation factor as defined by \citet{cowan1981}. 
                               In Cols\@. 3 and 6, e represents even and o odd.
                               This table is only available via the GAVO service TOSS (\url{http://dc.g-vo.org/TOSS}).} \\
\hline\hline
\noalign{\smallskip}
                 & \multicolumn{3}{c}{Lower level} && \multicolumn{3}{c}{Upper level} &           &      &    \\
\cline{2-4}\cline{6-8}
\noalign{\smallskip}
Wavelength{$^a$} / \AA & Energy{$^b$} / cm$^{-1}$ & Parity & $J$  && Energy{$^b$} / cm$^{-1}$ & Parity & $J$  & $\log g_\mathrm{i}f_\mathrm{ik}$ & $g_\mathrm{k}A_\mathrm{ki}$ / s$^{-1}$  & CF \\
\noalign{\smallskip}
\hline
\endfirsthead
\caption{continued.}\\
\hline\hline
\noalign{\smallskip}
                 & \multicolumn{3}{c}{Lower level} && \multicolumn{3}{c}{Upper level} &           &      &    \\
\cline{2-4}\cline{6-8}
\noalign{\smallskip}
Wavelength{$^a$} / \AA & Energy{$^b$} / cm$^{-1}$ & Parity & $J$  && Energy{$^b$} / cm$^{-1}$ & Parity & $J$  & $\log g_\mathrm{i}f_\mathrm{ik}$ & $g_\mathrm{k}A_\mathrm{ki}$ / s$^{-1}$  & CF \\
\noalign{\smallskip}
\hline
\noalign{\smallskip}
\endhead
\hline
\noalign{\smallskip}
\endfoot
\noalign{\smallskip}
\hline
\noalign{\smallskip}
\multicolumn{11}{p{15.4cm}}{{Notes.\,
                             $^{(a)}$\,All wavelengths (given in vacuum for                         $\lambda <   2000\,\mathrm{\AA}$, 
                                                                   air for  $2000\,\mathrm{\AA} \le \lambda \le 20000\,\mathrm{\AA}$,
                                                                vacuum for $20000\,\mathrm{\AA} <   \lambda$) are deduced from experimental energy levels.}
                             $^{(b)}$\,Experimental energy levels taken from \citet{kramidaetal2019}}.\\
\end{longtable}

\begin{longtable}{rrccrrccrrr}
\caption{\label{tab:cuvii:loggf}Calculated HFR oscillator strengths ($\log g_\mathrm{i}f_\mathrm{ik}$) and transition probabilities ($g_\mathrm{k}A_\mathrm{ki}$) 
                               in \ion{Cu}{vii}. CF is the
                               absolute value of the cancellation factor as defined by \citet{cowan1981}. 
                               In Cols\@. 3 and 6, e represents even and o odd.
                               This table is only available via the GAVO service TOSS (\url{http://dc.g-vo.org/TOSS}).} \\
\hline\hline
\noalign{\smallskip}
                 & \multicolumn{3}{c}{Lower level} && \multicolumn{3}{c}{Upper level} &           &      &    \\
\cline{2-4}\cline{6-8}
\noalign{\smallskip}
Wavelength{$^a$} / \AA & Energy{$^b$} / cm$^{-1}$ & Parity & $J$  && Energy{$^b$} / cm$^{-1}$ & Parity & $J$  & $\log g_\mathrm{i}f_\mathrm{ik}$ & $g_\mathrm{k}A_\mathrm{ki}$ / s$^{-1}$  & CF \\
\noalign{\smallskip}
\hline
\endfirsthead
\caption{continued.}\\
\hline\hline
\noalign{\smallskip}
                 & \multicolumn{3}{c}{Lower level} && \multicolumn{3}{c}{Upper level} &           &      &    \\
\cline{2-4}\cline{6-8}
\noalign{\smallskip}
Wavelength{$^a$} / \AA & Energy{$^b$} / cm$^{-1}$ & Parity & $J$  && Energy{$^b$} / cm$^{-1}$ & Parity & $J$  & $\log g_\mathrm{i}f_\mathrm{ik}$ & $g_\mathrm{k}A_\mathrm{ki}$ / s$^{-1}$  & CF \\
\noalign{\smallskip}
\hline
\noalign{\smallskip}
\endhead
\hline
\noalign{\smallskip}
\endfoot
\noalign{\smallskip}
\hline
\noalign{\smallskip}
\multicolumn{11}{p{15.4cm}}{{Notes.\,
                             $^{(a)}$\,All wavelengths (given in vacuum for                         $\lambda <   2000\,\mathrm{\AA}$, 
                                                                   air for  $2000\,\mathrm{\AA} \le \lambda \le 20000\,\mathrm{\AA}$,
                                                                vacuum for $20000\,\mathrm{\AA} <   \lambda$) are deduced from experimental energy levels.}
                             $^{(b)}$\,Experimental energy levels taken from \citet{vanhethofetal1990}}.\\
\end{longtable}

\twocolumn
\end{appendix}

\end{document}